\documentclass{ifacconf}

\usepackage{graphicx}      % include this line if your document contains figures
\usepackage{natbib}        % required for bibliography

\usepackage[T1]{fontenc}%
\usepackage{amsmath}
\usepackage{amssymb}
\usepackage{amsfonts}
\usepackage{booktabs}
\usepackage{multirow}
\usepackage{xspace}
\usepackage{color}
\usepackage{wrapfig}
\usepackage{url}
\usepackage{stmaryrd}
\usepackage{comment}
\usepackage{mathops}
\usepackage{colortbl}
\usepackage[long]{optidef}
\usepackage{todonotes}

\usepackage[all]{xy}
\SelectTips{cm}{}

\usepackage{algorithm}
\usepackage[]{algpseudocode} 
\algnewcommand{\IIf}[1]{\State\algorithmicif\ #1\ \algorithmicthen}
\algnewcommand{\EndIIf}{\unskip\ %\algorithmicend\ \algorithmicif
}

\newcounter{researchquestionCount}
\newcommand{\researchquestion}[1]{\stepcounter{researchquestionCount}\vspace{10pt}\noindent{\bf RQ\arabic{researchquestionCount} {\it #1\hfill}}\vspace{4pt}}

\newcommand{\Model}{\mathcal{M}}

\newcommand{\DefSpec}{:\equiv}

\newcommand{\ATBRKIN}{\text{AT1}}
\newcommand{\ATBRKINca}{\BoxOp{[0,30]} (\mathsf{rpm} \le p)}
\newcommand{\ATBRKINcb}{\BoxOp{[0,30]} (\mathsf{speed} \le 60)}
\newcommand{\ATBRKINcc}{\DiaOp{[0,30]} (\mathsf{gear} \ge 3)}

\newcommand{\ATSpeed}{\text{AT2}}
\newcommand{\ATSpeedca}{ \DiaOp{[0,29]}(\mathsf{speed} \ge 100)}
\newcommand{\ATSpeedcb}{ \DiaOp{[29,30]}(\mathsf{speed} \le 65)}
\newcommand{\ATSpeedFormula}{ \ATSpeedca \land \ATSpeedcb }

\newcommand{\ATSpeedRpm}{\text{AT3}}
\newcommand{\ATSpeedRpmca}{ \DiaOp{[0,10]}(\mathsf{speed} \ge p_1)}
\newcommand{\ATSpeedRpmcb}{\BoxOp{[0,30]}(\mathsf{rpm} \le p_2)}
\newcommand{\ATSpeedRpmFormula}{ \ATSpeedRpmca \land  \ATSpeedRpmcb }
\newcommand{\AFCSpec}{\text{AFC}}
\newcommand{\AFCca}{\BoxOp{[31, 50]} (\mathsf{mode} = 0) }
\newcommand{\AFCcb}{\DiaOp{[11, 20]}(\mathsf{mode} = 1)}
\newcommand{\AFCcc}{\BoxOp{[0,30]}(\mathsf{throttle} > 40 \Rightarrow \mathsf{engine} < 1000)}
\newcommand{\AFCcd}{\DiaOp{[0, 50]}\BoxOp{[0,25]}(\mathsf{engine} > 1000)}

\newcommand{\WTSpec}{\text{WT}}
\newcommand{\WTca}{\DiaOp{[0,90]}\BoxOp{[0,5]}(15.5 \le v \le 15.95 \land \theta < 12)}
\newcommand{\WTcb}{\DiaOp{[0,90]} (M_{g,d} \ge 47000)}
\newcommand{\WTcc}{\DiaOp{[0, 90]}(\Omega < 9)}

\newcommand{\Robj}{\mathsf{RObj}}
\newcommand{\Rcon}{\mathsf{RCon}}
\newcommand{\Rnv}{\mathsf{RVNum}}
\newcommand{\Rob}{\mathsf{rb}}
\newcommand{\Pop}{\langle x_1, x_2, \dots, x_\lambda\rangle}
\newcommand{\Indiv}{x}

\theoremstyle{plain}

\newcommand{\envalias}[2]{\newenvironment{#1}{\begin{#2}}{\end{#2}}}
\envalias{mytheorem}{thm}
\envalias{myproposition}{prop}
\envalias{mylemma}{lem}
\envalias{myexample}{exmp}
\envalias{myremark}{rem}
\envalias{mydefinition}{defn}
\envalias{myproblem}{prob}

\SetSymbolFont{stmry}{bold}{U}{stmry}{m}{n}

\specialcomment{auxproof}
{\mbox{}\newline\textbf{BEGIN: AUX-PROOF}\dotfill\newline}
{\mbox{}\newline\textbf{END: AUX-PROOF}\dotfill\newline}
\excludecomment{auxproof}

\begin{document}
\begin{frontmatter}

\title{Constrained Optimization for Hybrid System Falsification and Application to Conjunctive Synthesis}
%
% \thanks[footnoteinfo]{Sponsor and financial support acknowledgment
% goes here. Paper titles should be written in uppercase and lowercase
% letters, not all uppercase.}

\author[NII,SOKENDAI]{Sota Sato} 
\author[KUIS,NII]{Masaki Waga} 
\author[NII,SOKENDAI]{Ichiro Hasuo}

\address[NII]{
National Institute of Informatics, Tokyo, Japan
\\(e-mail: \{sotasato,mwaga,hasuo\}@nii.ac.jp)}

\address[SOKENDAI]{The Graduate University for Advanced Studies (SOKENDAI), Tokyo, Japan}

\address[KUIS]{Graduate School of Informatics, Kyoto University, Japan}

% Abstract of not more than 250 words.
\begin{abstract}
  The \emph{synthesis} problem of a cyber-physical system (CPS) is
  to find an input signal under which the system's behavior
 satisfies a given specification. Our setting is that the specification is a formula of signal temporal logic, and furthermore, that the specification is a \emph{conjunction} of different and often conflicting requirements.  Conjunctive synthesis is often challenging for \emph{optimization-based falsification}---an established method for CPS analysis that can also be used for synthesis---since the usual framework (especially how its \emph{robust semantics} handles Boolean connectives) is not suited for finding delicate trade-offs between different requirements. Our proposed method consists of a combination of optimization-based falsification and \emph{constrained optimization}. Specifically, we show that the state-of-the-art \emph{multiple constraint ranking} method can be combined with falsification powered by CMA-ES optimization; its performance advantage for conjunctive synthesis is demonstrated in experiments.
% \mw{I added the phrase ``conjunctive synthesis'' in the last sentence.}

% we deal with specifications given by formulas of signal temporal logic; moreover, we focus on conjunctive specifications---requiring multiple constraints, often conflicting with each other, to be simulatneously satisfied. 

% usually described in terms of of \emph{signal temporal logic}.
%   Specifically, such a subclass of \emph{parameter synthesis} naturally appears in industry practice  that specification is characterized by a conjunction of multiple specifications, namely $\varphi_1 \land \dots \land \varphi_m$. This is what we call \emph{conjunctive synthesis}.
  % It is known that \emph{optimization-based falsification}, an established methodology in \emph{cyber-physical system}, can be applied to solving \emph{parameter synthesis}.
  % In optimization-based approach, however, there is a drawback of performance identified as \emph{scale problem} and latent inherently in \emph{conjunctive synthesis}.
  % This paper proposes an approach to \emph{conjunctive synthesis} that keeps away from \emph{scale problem} by adapting \emph{multiple constraint ranking}, a state-of-the-art algorithm in the field of \emph{constrained optimization}.
  % Our experiment shows that our approach indicates consistent performance even in critical cases such that existing methodology can hardly synthesize a satisfying input.
\end{abstract}

\end{frontmatter}
%===============================================================================
% \todo[inline]{We need to decide the keywords}
%####################
\section{Introduction}\label{sec:intro}
%####################

% \todo[inline]{Try other algorithms for constrained optimization and make comparison?}

% \paragraph{Optimization-Based Falsification}
\paragraph*{Hybrid System Falsification}

Quality assurance of cyber-physical systems (CPSs) is a problem of grave social, economic and humanitarian importance. Given the limited applicability of \emph{formal verification} to CPSs---due to fundamental/ challenges such as the inherent hybrid dynamics in CPSs and the absence of white-box system models---one would often turn to so-called \emph{light-weight formal methods}. \emph{(Hybrid system) falsification} is one such method that attracts attention from industry and academia alike; see e.g.,~\cite{ARCH20}. 

% \emph{Formal verification} is a quality assurance method commonly applied to software; its unique advantage is that it can give an absolute correctness guarantee for absence of errors, in the form of mathematical proofs. However, applicability of formal verification to CPSs is limited because of a few fundamental challenges. One challenge is the aspect of CPSs as \emph{hybrid systems}: the continuous notions of space and time make the configuration spaces infinite and therefore not amenable to exhaustive model checking. Another challenge is the common unavailability of white-box system models of  target CPSs: if formal verification is about proving safety as a ``theorem,'' then a white-box system model is a ``definition'' on which a proof is established. 

% \emph{Falsification} is a major approach to quality assurance of \emph{cyber-physical systems}.

The falsification problem is to discover a counterexample for a logically-formulated system specification. It is seen as one specific realization of \emph{search-based testing}. 
\begin{myproblem}[Falsification]\label{prob:Falsification}\hfill
     \begin{itemize}
    \item{\textbf{Given:}} 
      a \emph{model} $\Model$ that takes an input signal $u$
      and  yields an output signal $\Model(u)$, and
      a \emph{specification} $\varphi$ % (a temporal formula)
    \item{\textbf{Find:}} 
      a \emph{falsifying input}, that is, an input signal $u$ such
      that the corresponding output $\Model(u)$ violates $\varphi$ 
    \end{itemize}
\end{myproblem}
Here the system model $\Model$ is typically that of a hybrid system (car, airplane, etc.) and is often given in the form of a Simulink model.

 \emph{Optimization-based falsification}, initiated
in~\cite{FainekosP09}, is an established approach to the falsification
problem.  The key idea of optimization-based falsification is
the translation of Problem~\ref{prob:Falsification} into minimization of
the robustness with which $\Model(u)$ satisfies the specification $\varphi$.  As
we will see in \S{}\ref{sec:byConstrainedOptim}, such a degree is
incarnated by the \emph{robust semantics}, which assigns an (extended)
real number $\sem{\Model(u),\varphi} \in \Rextended$ to an output signal
$\Model(u)$ and a specification $\varphi$.  This allows us to utilize
existing optimization algorithms to adaptively choose prospective input
signals $u$ and eventually find a falsifying input. 

Note that, in optimization-based falsification, the model $\Model$ in Problem~\ref{prob:Falsification} need not be a white-box one. One does not need to understand its internal working; it is enough to be able to observe the output $\Model(u)$ that corresponds to a given input $u$. 

\paragraph*{Conjunctive Synthesis via Hybrid System Falsification}
One easily figures out that falsification of $\lnot\varphi$ (i.e., to find a falsifying input for $\lnot\varphi$)
is equivalent to \emph{synthesis} for $\varphi$ (i.e., to find an input that satisfies $\varphi$).  
%
% \begin{wrapfigure}[5]{r}{0.3\textwidth}
% \vspace{-2.3em}
% \begin{math}
%    	\xymatrix@1@+0.8em{
%   	 {}
%    	  \ar[r]^-{u}
% 	   &
% 	   { \hspace{-0.8em} \quad\xybox{ *+++++++[F]{\Model} }}
% 	     \ar[r]^-{\Model(u)}_-{%\not
%                                         \models\varphi \; ?}
% 		   &
%  	  {}
% 	   }
%   \end{math}
% \caption{The conjunctive synthesis problem, where $\varphi\equiv\varphi_1 \land  \dots \land \varphi_m$}
% \end{wrapfigure}
%
%\paragraph{Conjunctive Synthesis}
The goal of the paper is to exploit and extend the  techniques for optimization-based falsification with the aim of solving the following \emph{conjunctive synthesis problem}. 
\begin{myproblem}[Conjunctive Synthesis]\label{prob:ConjSynth}\hfill
     \begin{itemize}
    \item{\textbf{Given:}} 
      a \emph{model} $\Model$ that takes an input signal $u$
      and  yields an output signal $\Model(u)$, and
      a \emph{conjunctive specification} $\varphi\equiv \varphi_1 \land \varphi_2 \land \dots \land \varphi_m$ 
% (a temporal formula)
    \item{\textbf{Find:}} 
      a \emph{satisfying input}, that is, an input signal $u$ such
      that the corresponding output $\Model(u)$ satisfies $\varphi$ 
    \end{itemize}
\end{myproblem}

Instances of conjunctive synthesis are omniscient in the real-world system design processes. 
\begin{myexample}\label{ex:leading}
Here is our leading example. The system model $\Model$ is given by a Simulink model for automatic transmission system---a system model commonly used in the falsification literature such as~\cite{ARCH20}. The model $\Model$ has two inputs ($\mathsf{throttle}, \mathsf{brake}$) and three outputs ($\mathsf{rpm}, \mathsf{speed}, \mathsf{gear}$). The specification $\varphi$ is given by
 \begin{align}\label{eq:leadingSpec}
\begin{aligned}
  \ATBRKIN_p \DefSpec \enspace & \BoxOp{[0,30]} (\mathsf{rpm} \le p) \\
 & \land \BoxOp{[0,30]} (\mathsf{speed} \le 60) \\
 & \land \DiaOp{[0,30]} (\mathsf{gear} \ge 3);
\end{aligned} 
\end{align}
it means the gear should reach the third without any of RPM and speed getting too large, a requirement common in the break-in procedure. The RPM bound $p$ is a parameter: the smaller $p$ is, the harder the synthesis problem becomes. 
\end{myexample}
Conjunctive synthesis tends to be hard when it contains conflicting requirements.
This is the case with  the above leading example---gear gets larger typically when RPM and/or speed is larger, while the specification $\varphi$ requires RPM and speed to stay small.
In this situation, satisfying all requirements asks for a careful trade-off between them.
Automated synthesis of input signals that achieve such delicate trade-offs (between performance and energy-efficiency, safety and progress, etc.) can help system designers who would otherwise spend a lot of time for manual trials and search.

\paragraph*{The Scale Problem}
Turning now from the problem itself to a solver for conjunctive synthesis,
a falsification solver can be used as an approach since Problem~\ref{prob:ConjSynth} is equivalent to falsification of $\lnot\varphi$ (Problem~\ref{prob:Falsification}).
However, existing falsification solvers can struggle with some problem instances, as we will see in~\S{}\ref{sec:expr}. The main obstacle is the \emph{scale problem}, a general problem in falsification that is identified and tackled in~\cite{Zhang2019}. 

To illustrate the scale problem, consider our leading example where the specification $\ATBRKIN_p
% \equiv \ATBRKINFormula
$ is given in~(\ref{eq:leadingSpec}).
Following the standard definition by~\cite{FainekosP09} (see \S{}\ref{sec:byConstrainedOptim}), the robust semantics $\sem{\Model(u), \ATBRKIN_p} \in \Rextended$ of the formula $\ATBRKIN_p$ is given by
\begin{align}
\begin{aligned}
 &\sem{\Model(u), \ATBRKIN_p}=v_{1}\sqcap v_{2}\sqcap v_{3},\quad\text{where}
 \\
    & v_{1}=\Wedge{t \in [0,30]}\bigl(p - \mathsf{rpm}(t)\bigr), \quad
      v_{2} =\Wedge{t \in [0,30]}\bigl(60 -\mathsf{speed}(t)\bigr), \\
    & v_{3}=\Vee{t \in [0,30]}\bigl(\mathsf{gear}(t) - 3\bigr). 
\end{aligned}    
\label{eq:ATBRKINsem}
\end{align}
This expression clearly shows that, in order for $\Model(u)$ to satisfy the specification $\ATBRKIN_p$ (which is equivalent to $\sem{\Model(u), \ATBRKIN_p}>0$), 
all  the three values $v_{1},v_{2}, v_{3}$
% of the three components
composed with infimum $\sqcap$ 
% (namely $v_{1}:=\Wedge{t \in [0,30]}\bigl(p - \mathsf{rpm}(t)\bigr)$, 
% $v_{2}:=\Wedge{t \in [0,30]}\bigl(60 -\mathsf{speed}(t)\bigr)$ and
% $v_{3}:=\Vee{t \in [0,30]}\bigl(\mathsf{gear}(t) - 3\bigr)$)
should be positive simultaneously.

The issue here is that the formulation~(\ref{eq:ATBRKINsem}) can prevent hill-climbing optimization algorithms from effectively making all the three values $v_{1},v_{2}, v_{3}$ positive.
% finding a solution
% with respect to each condition.
For example, imagine that we are in an early stage of the search for a satisfying input signal, and that all the values  $v_{1},v_{2}, v_{3}$ are still negative. 
By the nature of the automatic transmission model $\Model$,
the value $v_{1}$ is likely to be in the order of (minus) thousands,
the value $v_{2}$ is likely to be in the order of (minus) tens, and
the  value $v_{3}$ is either  $-2$ or $-1$. In this case, the value $v_{1}$ dominates the overall objective function $v_{1}\sqcap v_{2}\sqcap v_{3}$, masking the contribution of $v_{2}$ and $v_{3}$. This obviously does not help to find a delicate balance between $v_{1}, v_{2}$ and $v_{3}$.

% ranges in the order of (minus) thousands,
% the  value $v_{2}$ ranges in the order of (minus) tens, and
% %on the other hand
%  the  value $v_{3}$ ranges from $-$2 to 1.

% This means that the hill-climbing of the objective function $\sem{\Model(u), \ATBRKIN_p}$
% rarely takes account of the contribution of the third component,
% masked by the value of the other components.

This problem---that some specific component dominates the robustness of a Boolean combination and masks away the other components---is exhibited in~\cite{Zhang2019} where it is called the \emph{scale problem}. 
% As exhibited by~\cite{Zhang2019}, Boolean connectives pose the challenge of the \emph{scale problem}.
A solution is proposed in~\cite{Zhang2019} where conjuncts or disjuncts are thought of as arms in the \emph{multi-armed bandit problem} (MAB) and an MAB algorithm is combined with a hill-climbing optimization solver. However, this solution in~\cite{Zhang2019} is
% However, the MAB-based methods presented in~\cite{Zhang2019}
% are
 dedicated to satisfying the specifications of the form
$\DiaOp{I} (\varphi_1 \lor \varphi_2)$ and
$\DiaOp{I} (\varphi_1 \land \varphi_2)$; therefore it does not apply to our current problem of conjunctive synthesis (satisfying $\varphi_1 \land \varphi_2 \land \dots \land \varphi_m$, Problem~\ref{prob:ConjSynth}). 
 % case of satisfying
%  $\BoxOp{} \varphi_1 \lor \BoxOp{} \varphi_2$.

% falsify  
% falsify  $\BoxOp{} (\varphi_1 \land \varphi_2)$,

\paragraph*{Contribution}
% \mw{Here, the relationship between constrainted optimization and the conjunctive synthesis is implicit (and I do not think it is obvious). Giving an illustraiton will be helpful (if we have enough space).}
Our proposed method for solving conjunctive synthesis (Problem~\ref{prob:ConjSynth}) combines optimization-based falsification and \emph{constrained optimization}---optimization of the value of a single objective function but subject to potentially multiple constraints. Specifically, we show that the state-of-the-art \emph{multiple constraint ranking}  (MCR) method by~\cite{dePaulaGarcia2017} can be combined with \emph{CMA-ES}, an optimization algorithm by~\cite{Hansen2005} that is commonly used in optimization-based falsification. 

In the original work by~\cite{dePaulaGarcia2017}, MCR was introduced as a constraint handling technique (CHT) associated with the canonical genetic algorithm, while it was suggested that MCR could also be associated with other evolutionary algorithms. In this work, we integrate MCR with CMA-ES.
%for solving constrained optimization problems, and suggested it can be associated with other evolutionary algorithms, which leads us to combine MCR with CMA-ES.
 MCR address the scale problem in the  constraints and/or the objective function, without requiring user-specified parameters or additional computation to estimate scales. The key idea that allows MCR to do so is the use of suitable \emph{rankings}---instead of robustness values of constraints themselves---in prioritizing candidate solutions. 

We conduct an experimental evaluation of the proposed method, where we use hybrid system models from the falsification literature (such as~\cite{ARCH20}) and conjunctive specifications for synthesis. The comparison is against the state-of-the-art falsification tool Breach presented in~\cite{Donze10} (ver.\ 1.7.0, from January 2020); the experimental results show that our integration of MCR successfully addresses the scale problem and succeeds in conjunctive synthesis more often.

% MCR is proposed with the key advantage out of other CHT, that is, it overcomes the scale problem in the constraints and/or objective function, neither requiring user-specified parameters nor additional computation to estimate scales; instead it prioritizes the candidate solutions by building multiple \emph{rankings} with respect to the objective function and each constraint.
% Our experimental results indicate clear performance advantage over an existing falsification solver (we compare with Breach~\cite{Donze10}).

% \mw{add a summary of the contributions?}
\paragraph*{Related Work}

% Formal verification literature
There are various techniques toward formal verification of hybrid systems such as reachability analysis and theorem proving; see e.g.~\cite{ChenAS13,FrehseGDCRLRGDM11,FanQM0D16,DBLP:books/sp/Platzer18}. 
However, they are currently not very successful in the quality assurance of the \emph{real-world} CPSs due to scalability issues and the scarcity of  white-box system models.

% % Application to synthesis
% As a light-weight quality assurance method of CPSs, falsification attracts attention from both industry and academia. See, e.g.,~\cite{ARCH20}.
% Falsification is directly applicable to the synthesis problem (\cite{Donze2010}). 

% treatment of scale problem
The scale problem in falsification has been reported and several countermeasures have been proposed.
In~\cite{DokhanchiYHF17} and~\cite{FerrereNDIK19}, the scale problem is addressed by adapting the scales of robustness of subformulas. Moreover, the scale adaptation is automatic. 
In~\cite{Zhang2019}, an MAB-based algorithm is proposed: its hierarchical framework considers Boolean subformulas as arms in the multi-armed bandit problem, thus avoiding  superposition of robustness values (and in particular the scale problem) altogether. In~\cite{Zhang2019}, the limit of effectivity of  scale adaptation is experimentally shown, too. We note that the technique in~\cite{Zhang2019} does not apply to conjunctive synthesis---as we discussed in the above.

% but their method works only for a subclass of STL, which is strictly smaller than ours.

% perhaps \cite{SakamotoA19}?

\emph{Input constraints}---constraints that a falsifying input signal should satisfy---are important in real-world application of falsification. Efficient handling of input constraints has been studied in some recent works including~\cite{DBLP:conf/nfm/ZhangAH20,DBLP:conf/nfm/BarbotBDDKY20,DBLP:journals/tcad/ZhangAH20}. Note that the ``constraints'' in this paper are mostly on output signals, instead of on input.

% Besides, the term ``constraint'' in falsification often appears to constrain input signal for example in~\cite{DBLP:conf/nfm/ZhangAH20} and~\cite{DBLP:conf/nfm/BarbotBDDKY20}.

% constraint handling methods for CMA-ES
Finally, besides MCR from~\cite{dePaulaGarcia2017}, other constraint handling techniques that 
%can handle simulation-based constraints and
 go along well with CMA-ES are pursued in~\cite{ho_evolutionary_2007,spettel_multi-recombinative_2019,SakamotoA19}. They differ in target constraint classes; see \S{}\ref{subsec:taxonomy}.

% We note that constraint handling in CMA-ES is previously pursued in~\cite{SakamotoA19}. Their focus is however on a priori constraints, instead of simulation-based constraints that are our current setting.

\paragraph*{Organization of the Paper}
\S{}\ref{sec:byConstrainedOptim} introduces some background of optimization-based falsification and give the formalism of conjunctive synthesis by constraint optimization.
\S{}\ref{sec:ourAlgorithm} introduces our algorithm to solve constrained optimization.
\S{}\ref{sec:expr} reports the experiments for assessing the performance advantage of our approach,
and \S{}\ref{sec:conclusion} concludes the paper.

%####################
\section{Conjunctive Synthesis by Constrained Optimization}\label{sec:byConstrainedOptim}
%####################

\subsection{Boolean Semantics and Robust Semantics of STL}

% We start with defining signals and the syntax for STL.
Our specifications $\varphi\equiv \varphi_1 \land \dots \land \varphi_m$ are given by \emph{signal temporal logic} (STL) formula, as is common in the falsification literature.
  We fix the finite set $\Var$ of variables.

\begin{mydefinition}[signal]
  A \emph{signal} over $\Var$ is a function $\sigma: \Rnn \to \R^{\Var}$.
  A signal is called \emph{$M$-dimensional} when the cardinality of $\Var$ is $M$.
\end{mydefinition}

\begin{mydefinition}[STL syntax]\label{def:syntax}
  In STL, the set $\Fml$ of \emph{formulas} is defined as follows.
  \begin{equation}
    \begin{array}{rrl}
      \Fml \ni 
      &\varphi \,:=\,
      & \top
        \mid \bot
        \mid f(\vec{x}) > 0
        \mid f(\vec{x}) < 0
        \mid \neg \varphi 
        \mid \varphi \vee \varphi  \\
        && \mid \varphi \wedge \varphi
        \mid \varphi \UntilOp{I} \varphi 
        \mid \varphi \Release{I} \varphi 
        \mid \DiaOp{I} \varphi 
        \mid \BoxOp{I} \varphi 
        % \quad \text{ where } l \in \AP\\
    \end{array}
  \end{equation}
  Here $f$ is a function symbol applied to a tuple $\vec{x}$ of variables from $\Var$, and $f(\vec{x})>0$ is an atomic formula. An example of such $f$ is $f(x,y)= 4x - y - 3$; it denotes a function $\mathbb{R}^{n}\to \mathbb{R}$ for a suitable arity $n$.  $I$ is a non-singular interval in $\Rnn$,
  i.e., $I=[a,b]$, $(a,b)$, $(a,b]$, $[a,b)$ or $[a, \infty)$ where $a<b$.
\end{mydefinition}

In the conventional \emph{Boolean} semantics of STL,
a signal $\sigma$'s satisfaction of a formula $\varphi$ is
given by a relation $ \sigma \models \varphi$ (see~\cite{MalerN04}).
The Boolean semantics is not suited for optimization-based falsification,
which typically relies on hill-climbing optimization algorithms.
% Instead of this, we employ the \emph{robust semantics} of STL.

% Their \emph{Boolean semantics} $u\models\varphi$ is defined in a usual manner---it is much like for LTL formulas but takes additional care of the continuous notion of time. 

The \emph{robust semantics} $\sem{\sigma,\varphi}$ of an STL formula $\varphi$ under a signal $\sigma$ is the key enabler of optimization-based falsification; it is introduced in~\cite{DBLP:conf/fates/FainekosP06}. The robust semantics takes an (extended) real number as its value ($\sem{\sigma,\varphi}\in \Rextended$). The value
 designates how robustly the formula is satisfied.

\begin{mydefinition}[robust semantics]\label{def:robustSem}
  Let $\sigma \colon \Rnn \to \R^\Var$ be a signal 
  and $\varphi$ be a STL formula.
  We define the \emph{robustness} 
  $\Robust{\sigma}{\varphi} \in \Rextended$
  by induction, as shown in Table~\ref{table:robustness}. 
  Here $\sqcap$ and $\sqcup$ denote infimums and supremums of (extended) real numbers, respectively, 
  and $f(\sigma(0)(\vec{x}))$ denotes $ f(\sigma(0)(x_1), \sigma(0)(x_2),\dots,\sigma(0)(x_n))$.

  \begin{table*}[tbp]
   % \hspace{-4em}
   \centering
    \begin{tabular}{l}
      \scalebox{1}{
      \begin{math}
        \begin{array}{rll}
        \Robust{\sigma}{\top} & \Defeq & \infty\\
          \Robust{\sigma}{\bot} & \Defeq & 0\\
          \Robust{\sigma}{f(\vec{x}) > 0} & \Defeq & f(\sigma(0)(\vec{x}))\\
          \Robust{\sigma}{f(\vec{x}) < 0} & \Defeq & - f(\sigma(0)(\vec{x}))\\
          \Robust{\sigma}{\neg \varphi} & \Defeq & - \Robust{\sigma}{\varphi}\\
          % \Robust{\sigma}{\varphi_1 \vee \varphi_2} & \Defeq & 
          %                                                          \Robust{\sigma}{\varphi_1} \sqcup \Robust{\sigma}{\varphi_2}\\
        %   \Robust{\sigma}{\varphi_1 \wedge \varphi_2} & \Defeq & 
        % \Robust{\sigma}{\varphi_1} \sqcap \Robust{\sigma}{\varphi_2}\\
          \\
        \end{array}
        \quad
          \begin{array}{rll}
          \Robust{\sigma}{\varphi_1 \vee \varphi_2} & \Defeq & 
                                                                   \Robust{\sigma}{\varphi_1} \sqcup \Robust{\sigma}{\varphi_2}\\
          \Robust{\sigma}{\varphi_1 \wedge \varphi_2} & \Defeq & 
        \Robust{\sigma}{\varphi_1} \sqcap \Robust{\sigma}{\varphi_2}\\
          \Robust{\sigma}{\varphi_1 \UntilOp{I} \varphi_2}
            & \Defeq 
            & \Vee{t \in I} 
              (\Robust{\sigma^t}{\varphi_2} \sqcap 
              \Wedge{t' \in [0, t)} \Robust{\sigma^{t'}}{\varphi_1})\\
            \Robust{\sigma}{\varphi_1 \Release{I} \varphi_2}
            & \Defeq 
            & \Wedge{t \in I} 
              (\Robust{\sigma^t}{\varphi_2} \sqcup 
              \Vee{t' \in [0, t)} \Robust{\sigma^{t'}}{\varphi_1})\\
            % \Robust{\sigma}{\DiaOp{I} \varphi}
            % & \Defeq 
            % & \Wedge{t \in I} 
            %   (\Robust{\sigma^t}{\varphi}) \\
            % \Robust{\sigma}{\BoxOp{I} \varphi}
            % & \Defeq 
            % & \Vee{t \in I} 
            %   (\Robust{\sigma^t}{\varphi}) \\
        \end{array}
        \quad
          \begin{array}{rll}
          % \Robust{\sigma}{\varphi_1 \UntilOp{I} \varphi_2}
          %   & \Defeq 
          %   & \Vee{t \in I} 
          %     (\Robust{\sigma^t}{\varphi_2} \sqcap 
          %     \Wedge{t' \in [0, t)} \Robust{\sigma^{t'}}{\varphi_1})\\
          %   \Robust{\sigma}{\varphi_1 \Release{I} \varphi_2}
          %   & \Defeq 
          %   & \Wedge{t \in I} 
          %     (\Robust{\sigma^t}{\varphi_2} \sqcup 
          %     \Vee{t' \in [0, t)} \Robust{\sigma^{t'}}{\varphi_1})\\
            \Robust{\sigma}{\DiaOp{I} \varphi}
            & \Defeq 
            & \Wedge{t \in I} 
              (\Robust{\sigma^t}{\varphi}) \\
            \Robust{\sigma}{\BoxOp{I} \varphi}
            & \Defeq 
            & \Vee{t \in I} 
              (\Robust{\sigma^t}{\varphi}) \\
        \end{array}
    \end{math}
      }
 \end{tabular}
 \caption{Definition of robust semantics of STL}
 \label{table:robustness}
\end{table*}
\end{mydefinition}

The following relationship holds between the Boolean and robust semantics.

\begin{myproposition}\label{prop:robustAndBoolean}
    $\Robust{\sigma}{\varphi} > 0$ implies $\sigma \models \varphi$ and
    $\Robust{\sigma}{\varphi} < 0$ implies $\sigma \not\models \varphi$.
\end{myproposition}

\subsection{Conjunctive Synthesis by Constrained Optimization}
% \mw{Perhaps this subsection should not be in \S{}\ref{sec:byConstrainedOptim} to clearly distinguish what is known and our idea/observation.}
Noting that the robust semantics interprets conjunction $\land$ by the infimum $\sqcap$ of real numbers, Problem~\ref{prob:ConjSynth} can be solved as follows. 
% \begin{myproblem}[Conjunctive Synthesis by Optimization]
  % In the setting of Falsification, assume that the specification $\mathcal{S}$ is specified by an STL formula $\varphi$.
% In the setting of Problem~\ref{prob:ConjSynth}, 

Consider the following optimization problem:
 \begin{maxi}
  {u}{\sem{\Model(u), \varphi_1\land\dots\land\varphi_{m}},}{}{}
 \end{maxi} that is,
 \begin{maxi}
  {u}{\sem{\Model(u), \varphi_{1}}\sqcap\dots\sqcap\sem{\Model(u), \varphi_{m}}.}{}{\label{eq:ConjSynthByOptim}}
 \end{maxi}

If the discovered maximum is positive, then by Proposition \ref{prop:robustAndBoolean}, the corresponding value of the optimization variable $u$ is 
 % Once $u$ is found such that the above robustness value is positive, this $u$ is 
 a solution to conjunctive synthesis (Problem~\ref{prob:ConjSynth}). 
% This is the de Morgan dual of the very idea behind optimization-based falsification.

However, the form~(\ref{eq:ConjSynthByOptim}) that combines different values with infimum $\sqcap$ often hinders hill-climbing optimization---this is the scale problem that  we discussed in~\S{}\ref{sec:intro}. 

Our main idea in the paper is to regard conjuncts not as objectives (as in~(\ref{eq:ConjSynthByOptim})) but as constraints, described below.
\begin{myproblem}[Conjunctive Synthesis by Constrained Optim.]\label{prob:ConjSynthByMCO}
In the setting of Problem~\ref{prob:ConjSynth}, consider the following constrained optimization problem:
  % Assume that the specification $\mathcal{S}$ is specified by an STL formula $\varphi \equiv \varphi_{1}\lor\cdots\lor\varphi_{m}$.
 % Consider the following constrained optimization problem:
 \begin{maxi}
  {u}{[\![ \Model(u), \varphi_1 ]\!]}{}{\label{eq:constrained}}
  \addConstraint{[\![ \Model(u), \varphi_i ]\!]}{> 0, \quad}{i=2,\dots, m.}
 \end{maxi}
If the discovered maximum is positive, then by Proposition~\ref{prop:robustAndBoolean},  the corresponding value of the optimization variable $u$ is a solution to conjunctive synthesis (Problem~\ref{prob:ConjSynth}). 
\end{myproblem}

% Once $u$ is found such that the robustness $[\![ \Model(u), \varphi_1 ]\!]$ is positive, this $u$ is a solution to conjunctive synthesis (Problem~\ref{prob:ConjSynth}), by Proposition~\ref{prop:robustAndBoolean}. 
% Some remarks are in order.

\begin{myremark}\label{rem:choiceOfObjective}
In~(\ref{eq:constrained}), we picked $\varphi_{1}$ as the optimization target, leaving the other conjuncts $\varphi_{2},\dotsc,\varphi_{n}$ as constraints.
We can do so without loss of generality since the order of the objectives is irrelevant.
In practice, we would pick as $\varphi_{1}$ the conjunct that we expect to be the most challenging to satisfy. 
\end{myremark}

Via the translation of Problem~\ref{prob:ConjSynth} to Problem~\ref{prob:ConjSynthByMCO}, we make the major challenge in Problem~\ref{prob:ConjSynth} explicit in the problem formulation. Specifically, the conflict between different conjuncts $\varphi_{1},\dots,\varphi_{m}$ is buried away in a single objective function in~(\ref{eq:ConjSynthByOptim}); it is made explicit in~(\ref{eq:constrained}). However, this translation would lead to an efficient solution only if there exists an algorithm that successfully exploits the structure that is now made explicit. This is the topic of the next section (\S{}\ref{sec:ourAlgorithm}).

% \draftitem{
% \item spec: $\varphi_1 \lor \varphi_2 \lor \dots \varphi_m$ for any STL formula $ \varphi_1, \varphi_2, \dots, \varphi_m$
% \item AT model~\cite{ARCH20}
% \item Benchmark spec $\ATBRKIN_p \DefSpec \ATBRKINFormula$ with a ranging parameter $p$
% \item Goal is to find a falsifying signal
% \item Smaller the value of $p$ goes, the problem becomes harder to solve
% }

% \begin{myproblem}[falsification by optimization]\label{prob:falsificationByOptim}
%  In the setting of Falsification, assume that the specification $\mathcal{S}$ is specified by an STL formula $\varphi$.

% Consider the following optimization problem:
% \begin{displaymath}
%  \begin{array}{rl}
%   \min_{v}\;\sem{\Model(v), \varphi}. \\
%  \end{array}
% \end{displaymath}
%  Once $v$ is found such that $\sem{\Model(v), \varphi}<0$, we have $\Model(u,v)\not\models\varphi$ by boolean robust semantics
% \end{myproblem}

\begin{myremark}
We note that Problem~\ref{prob:ConjSynthByMCO} is a \emph{single-objective} optimization problem under multiple constraints, that should be distinguished from a \emph{multi-objective} optimization problem. The latter's notion of optimality is more involved (given e.g.\ by the \emph{Pareto front}), and it would call for very different algorithms. 
Recall that our ultimate goal is constraint satisfaction $\sem{\Model(u), \varphi_{1}}>0,\dotsc,\sem{\Model(u), \varphi_{m}}>0$. We need the robustness value of each formula $\varphi_{i}$ to be positive,
but  further optimization beyond $0$ is unnecessary.
\end{myremark}

\subsection{Optimization-Based Synthesis by CMA-ES}\label{sec:synthesisByCmaes}
For an optimization problem such as (\ref{eq:ConjSynthByOptim}), one can adopt any optimization algorithm on (extended) real function, as SA (simulated annealing), GNM (global Nelder-Mead), CMA-ES, and so on. 

Our algorithm for conjunctive synthesis (Problem~\ref{prob:ConjSynth}) will be based on CMA-ES, an evolutionary computation algorithm introduced by~\cite{Hansen2005}, whose efficiency in the context of optimization-based falsification is well-established~\cite{Zhang2019}.
Here is its outline.
\begin{mydefinition}[CMA-ES~\cite{Hansen2005}]\label{def:CMAES}
 Given a fitness function $f$,
 CMA-ES operates in an iterative manner,  repeating the following
 regenerational steps until termination.
 Let $\mu, \lambda$ be fixed natural numbers with $\mu < \lambda$.
 \begin{enumerate}
  \item Generate a \emph{population} $X = \langle u_1, u_2, \dots, u_\lambda\rangle$ by sampling from
        the distribution $d_{\theta}$ on the search space, with the parameter value $\theta$ that is previously chosen. 

	Specifically, the distribution $d_{\theta}$ is a Gaussian distribution, and $\theta$ gives its mean and covariance matrix.
  \item Select $\mu$ individuals
        $u_{1:\lambda}, u_{2:\lambda}, \dots, u_{\mu:\lambda} \in X$
        that are the fittest on $f$.
  \item Update the distribution parameter $\theta$ according to the selected individuals. 

	Specifically, the new mean is the mean of the selected individuals $u_{1:\lambda},  \dots, u_{\mu:\lambda}$, and the new covariance matrix is chosen in a suitable manner. See~\cite{Hansen2005}. 
 \end{enumerate}
\end{mydefinition}

Here we describe the algorithm of (pure) CMA-ES, in the form adapted to the current synthesis problem. In our algorithm later in~\S{}\ref{sec:ourAlgorithm}, it will be combined with MCR to address conflicting conjuncts in the specification.
% In this paper, we describe an algorithm for optimization-based synthesis integrated with CMA-ES.
Its pseudo code is presented in Algorithm~\ref{algo:SynthesisByCmaes}, in which three subroutines $\textsc{Ask}$, $\textsc{Tell}$, and $\textsc{Gen-Signal}$ appear.

\begin{algorithm}[hb]
\caption{CMA-ES adapted to optimization-based synthesis}
\label{algo:SynthesisByCmaes}
\begin{algorithmic}[1]
\Require a system model $\Model$, an STL formula $\varphi$, 
the size of population $\lambda$,
and an initial parameter $\theta^0$ of CMA-ES
\Function{Cmaes-Synthesize}{$\Model,\varphi, \lambda, \theta^0$}
%\State $\vec \bu \gets \vec \bu_0$
\State $\Rob \gets -\infty$\;; \quad $u \gets \bot$
\Statex
   \Comment{$\Rob$ is the greatest robustness so far with signal $u$}
\State $g \gets 0$
   \Comment{$g$ is generation of evolution}
\While{$\left(\parbox[]{4cm}{$\Rob \le 0$, within budget, and not stationary}\right)$}
  \label{line:whileGuard}
  \State $\Pop \gets \textsc{Ask}(\theta^g)$
  \Statex    \Comment{Sample population}

  \For{$i \gets 1$ to $\lambda$}
    \State $u_i \gets \textsc{Gen-Signal}(\Indiv_i)$
    \State $\Rob_i \gets \sem{\Model(u_i), \varphi}$
    \IIf{$\Rob_i > \Rob$}
      $\Rob \gets \Rob_{i}$\;; $u \gets u_i$ \EndIIf
      \Statex\Comment {Update robustness and signal}
  \EndFor
  \State \parbox[t]{7cm}{$\Indiv_1, \dots \Indiv_\lambda \gets \Indiv_{s(1)}, \dots, \Indiv_{s(\lambda)}$ \newline
         with $s(i) \Defeq \text{argsort}(\langle -\Rob_1, \dots, -\Rob_\lambda\rangle, i)$}
  \Statex  \Comment {\parbox[t]{6cm}{Sort individuals in descending order by their robustness}}
	\State $\theta^{g+1}\gets \textsc{Tell}(\theta^g, \Pop)$
     \Statex \Comment {Update parameter}
  \State $g\gets g+1$
\EndWhile%{$R\ge 0$ and within the budget}

\vspace{.3em}
\State$u\gets
\begin{cases}
 u & \text{if $\Rob>0$, that is, $\Rob=\sem{\Model(u), \varphi}>0$} \\
 \text{Failure}  & \text{no satisfying input found}
\end{cases}
$
\State \Return{$u$}
\EndFunction
\end{algorithmic}
\end{algorithm}

$\textsc{Ask}$ and $\textsc{Tell}$ are interface pattern of CMA-ES, encapsulating its detailed behavior; described in detail in ~\cite{collette_object-oriented_2010}.
Optimizers can be seen as a piece of program in the loop that receives evaluation results and generates new candidates to be evaluated next.
$\textsc{Ask}$ returns $\lambda$ sampled point $\Pop$ by parameter $\theta$ and $\textsc{Tell}$ updates the parameter by receiving sorted points $\langle\Indiv_{s(1)}, \dots, \Indiv_{s(\lambda)}\rangle$.

$\textsc{Gen-Signal}$ is a function that generates an input signal $u_{i}$ from an individual $x_{i}$. As usual with hybrid system falsification, we restrict input signals to piecewise constant ones, and represent them by the values of variables at each control point. Therefore our individuals $x_{i}$ resides in the set $\R^{m\cdot |\Var|}$, where $m$ is the number of control points and $\Var$ is the set of input variables to the model $\Model$.

% mediates a $x_{i}\in\R^{n}$ and sampling signals from $\R^\Var$.
% Since CMA-ES (and most optimization algorithm) cannot handle an infinite dimensional space $\R^\Var$ as a search space, in practical cases one can restrict the search space into the subspace of $\R^\Var$, represented by finite dimensional space $\R^n$.
% Typical solution is \emph{piecewise constant signal}.
% approximate signal
% that is sufficient to find satisfying iunput.
% In usual practical implementation of optimization-based falsification, like Breach~\cite{Donze10}, has such a mechanism.

On the guard for the while loop (Line~\ref{line:whileGuard}), we terminate if one of the following is the case: 1) $\Rob > 0$ (in which case $u$ is an input signal we look for); 2) the budget is used up (the budget can be specified by time or by the number of iterations); and 3) CMA-ES is in \emph{termination criteria}, numeric stationarity conditions   described in~\cite{Hansen2005}. 
% in the sense that the algorithm is in numerical stability that may become a waste of CPU-time to continue. \todo{Say ``stationary''!}
For instance, a termination criterion \texttt{EqualFunValues} is true if the best objective function values does not change in the last $10 + \lceil 30n / \lambda \rceil$ generations, where $n$ is the dimension of the search space and $\lambda$ is the size of population.
% Further criteria are described in~\cite{Hansen2005}.

% Obviously, $\textsc{Cmaes-Synthesize}(\Model, \lnot\varphi, \lambda, \theta^g)$ can play a role of falsification of $\varphi$.

%####################
\section{Our Algorithm:  Combining MCR and CMA-ES}\label{sec:ourAlgorithm}
%####################
\subsection{ The Constrained Optimization Taxonomy}\label{subsec:taxonomy}

% \mw{Move to the related work section?}
In search of a disciplined method for handling multiple (possibly conflicting) constraints, we turned to the taxonomy of constrained optimization problems~\cite{digabel2015taxonomy}. According to the taxonomy, our current problem 
 (Problem~\ref{prob:ConjSynthByMCO}) is classified as follows. 
\begin{itemize}
 \item \emph{Simulation-based} as opposed to \emph{a priori}, in the sense that the satisfaction of a constraint is determined only in a black-box manner (we assume a model $\Model$ is complex and thus black-box).
 \item \emph{Relaxable} as opposed to \emph{unrelaxable}, in the sense that the objective function has well-defined values even if the constraints are not satisfied.
 \item \emph{Quantifiable} as opposed to \emph{nonquantifiable}, in the sense that the degree of satisfaction of each constraint can be quantified (namely by the robustness $[\![ \Model(u), \varphi_i ]\!]$).
\end{itemize}

We note that constraint handling in CMA-ES is previously pursued in~\cite{SakamotoA19}. Their focus is however on a priori constraints, instead of simulation-based constraints that are our current setting.

\subsection{MCR in CMA-ES}
It turns out that the \emph{multiple constrained ranking} algorithm  (MCR)~\cite{dePaulaGarcia2017} fits the very classification discussed in the above. It works with general evolutionary optimization algorithms, hence also with CMA-ES.

\begin{mydefinition}[MCR in CMA-ES]\label{def:MCRInCMAES}
Assume the setting of Problem~\ref{prob:ConjSynthByMCO}, and let us write $f(u)=[\![ \Model(u), \varphi_1 ]\!]$ for the optimization objective (the fitness function). 

MCR in CMA-ES consists of replacing the use of the fitness function $f$, in Step~2 of Def.~\ref{def:CMAES}, with the following \emph{scoring function} $F_{X}$. (Note that the function $F_{X}$ relies on the current population $X$.)

For each individual $u$, the value $F_{X}(u)$ is a natural number, and those $u$ with smaller $F_{X}(u)$ are deemed fitter. Its value is defined by
\begin{equation} \label{eq:rankingFunction}
   F_{X}(u) \Defeq \begin{cases}
       \Rnv_{X}(u) + \sum_{j=2}^{m} \Rcon_{X}^{j}(u)\\
	       \qquad\text{(if no \emph{feasible solution} in population $X$)} \\ 
	       % & \text{in the sense that $[\![ \Model(u), \varphi_i ]\!] > 0$ (for all $i=2,\dots, m$) holds for no $u\in X$}\\
       \Robj_{X}(u) + \Rnv_{X}(u) + \sum_{j=2}^{m} \Rcon^{j}_{X}(u) \\
         \qquad \text{(otherwise)} \\
   \end{cases} 
\end{equation}
Here a feasible solution is an individual $u$ that satisfies all the constraints, that is,  $[\![ \Model(u), \varphi_i ]\!] > 0$ for all $i=2,\dots, m$. The functions $\Rnv_{X}, \Robj_{X}, \Rcon^{j}_{X}$ all return a suitable ``rank'' of the input. Specifically, 
\begin{itemize}
 \item $\Robj_{X}(u)$ is the rank of $u$ among the population $X$ in the descending order of the value of the objective $f$, 
      %That is\mw{``For example'' instead of ``That is'' ?}, $\Robj_{X}(u_{1})=1$ if $f(u_{1})$ is the greatest among $\{f(u)\mid u\in X\}$.
       i.e., $\Robj_{X}(u) = 1 + |\{u' \in X \mid f(u) < f(u')\}|$.
 \item For each $j\in [2,m]$, $\Rcon^{j}_{X}(u)$ is the rank of $u$ among the population $X$ in the descending order of the value of $0\sqcap [\![ \Model(\place), \varphi_j ]\!]$, that is, the opposite of the degree of violation of constraint $\varphi_{j}$. This is much like $\Robj_{X}(u)$ but note that the degree of satisfaction (i.e., positive robustness) is disregarded via $0\sqcap$.
 \item $\Rnv_{X}(u)$ is the rank of $u$ among $X$ in the \emph{ascending} order of the number of violated constraints, that is, for how many $j\in[2,m]$ we have $[\![ \Model(u), \varphi_j ]\!] \le 0$. 
\end{itemize}

\end{mydefinition}

% For optimization problem without constraints,
% one can identify its objective function with fitness function $f$
% to solve that optimization problem with CMA-ES.
% MCR replaces a fitness function with the following \emph{scoring function} $F(v)$
% to adapt this regenerational steps with constrained optimization problem~(\ref{eq:constrained}).

% A \emph{scoring function} $F(v)$ is a function which gives a real value to each individual $v$ among population.
% \begin{equation}
%    F(v) = \begin{cases}
%        \Rnv + \sum_{j=2}^{m} R_{\varphi_j} & \text{if no feasible solution in population $X$} \\ 
%        R_{f} + \Rnv + \sum_{j=2}^{m} R_{\varphi_j} & \text{otherwise} \\
%    \end{cases} 
% \end{equation}
% In MCR, the scoring function employs three kinds of ranks $R_f$, $R_{\varphi_i}$ and $\Rnv$.
% $R_f$ compares the value of the objective function $f$, namely
% the value of $[\![ \Model(v), \varphi_1 ]\!]$
% in the setting of the problem~(\ref{eq:constrained}).
% $R_{\varphi_i}$ ($i=2,\dots,m$) compares the violation of $\varphi_i$, namely
% $0 \lor [\![\Model(v), \varphi_i]\!]$.
% $\Rnv$ compares the number of violated constraints.
% Note that the value of $F(v)$ depends on population $X$.

 % figure, ranking table 
 \begin{table}[tb]
  \begin{center}
    \caption{Example; robustness of  $\varphi_1, \varphi_2, \varphi_3$ for each individual and their infimum} \label{tab:mcrRob}
    \begin{tabular}{l|rrr|r}
        Individual &$\varphi_1$&$\varphi_2$&$\varphi_3$& infimum\\ \hline
             $u_1$ &      1400 &      59.9 &      $-$2 & $-$2\\
             $u_2$ &      $-$9 &         2 &         1 & $-9$\\
             $u_3$ &    $-$180 &         2 &      $-$1 & $-$180\\
      \end{tabular}
  \end{center}
 \end{table}
 \begin{table}[tb]
  \begin{center}
    \caption{Example; values of ranks and scoring function for each individual} \label{tab:mcrRank}
    \begin{tabular}{l|rrrr|r}
        Individual & $\Robj_{X}$& $\Rcon_X^2$& $\Rcon_X^3$& $\Rnv_X$ & $F_X$ \\ \hline
             $u_1$ &         1 &         1 &         3 &   2 & 7 \\
             $u_2$ &         2 &         1 &         1 &   1 & 5 \\
             $u_3$ &         3 &         1 &         2 &   2 & 8 \\
      \end{tabular}
  \end{center}
 \end{table}
\begin{myexample} \label{ex:AT1withMCR}
Let $\varphi_1 \DefSpec \BoxOp{[0,30]} (\mathsf{rpm} \le 2400)$, $\varphi_2 \DefSpec \BoxOp{[0,30]} (\mathsf{speed} \le 60)$, $\varphi_3 \DefSpec \DiaOp{[0,30]} (\mathsf{gear} \ge 3)$ in the setting of Example~\ref{ex:leading}. For certain individuals $u_1, u_2, u_3$, their robustness values  $\sem{\Model(u_i), \varphi_j}$ with respect to $\varphi_{1},\varphi_{2}, \varphi_{3}$ are shown in Table~\ref{tab:mcrRob}.

% assume that we have a population $X = \langle u_1, u_2, u_3 \rangle$ where
% the robustness $\sem{\Model(u_i), \varphi_j}$ ($i,j=1,2,3$) is obtained as described in Table~\ref{tab:mcrRob}.

Assume that the population $X$ is $X=\langle u_1, u_2, u_3 \rangle$. 
In this case, the usual robust semantics indicates the input $u_1$ is the best individual among $X$, in the sense that the infimum is the largest. However, once we inspect the input signals $u_1, u_2, u_3$, it becomes obvious that $u_{1}$ is the farthest from desired---it is in fact the signal in which $\mathsf{brake}$ is constantly the maximum and $\mathsf{throttle}$ is constantly $0$.

% However, the input $u_{1}$ is in fact one in which $\mathsf{brake}$ is the maximum and $\mathsf{throttle}$ is $0$ all the time. This is far from the satisfying input signal that we are after, but it is deemed to be the best among $u_{1}, u_{2}, u_{3}$ because the scale of the robustness of $\varphi_{3}$ is small (the scale problem). \mw{It seems the signals $u_1, u_2, u_3$ are not shown, right? It is nice if they are illustrated e.g. by a figure.}

% can be a scenario of just pressing the accelerator intensively, which is unrelated with our goal.

The last mismatch between the robustness-based preference and human intuition comes from the scale problem. The quantity in $\varphi_{3}$ (namely $\mathsf{gear}$) is smaller in scale compared to those in $\varphi_{1}$ and $\varphi_{2}$; and therefore the robustness of $\varphi_{3}$ tends to mask that of others.

In contrast, the  MCR scoring function  gives different preference, as shown in Table~\ref{tab:mcrRank}.
Here we pick the formula  $\varphi_1 \equiv \BoxOp{[0,30]} (\mathsf{rpm} \le p)$ as the objective (Problem~\ref{prob:ConjSynthByMCO} and Definition~\ref{def:MCRInCMAES}); the others $\varphi_2, \varphi_3$ are deemed to be as constraints.

The scoring function $F_X$ indicates the best input is $u_2$. This matches human intuition: the input signal $u_{2}$ is one with moderate throttle and no braking. The signal $u_{2}$ satisfies $\varphi_2, \varphi_3$ and almost satisfies $\varphi_{1}$, violating the RPM limit $2400$ only by $9$.

% which keeps the gear low and modestly violates $\varphi_1$.
% In light of our requirement which asks for a careful trade-off between $\varphi_1, \varphi_2, \varphi_3$,
% it is natural to expect $u_2$ is more promising than $u_1$.

\begin{auxproof}
(** This is speaking about a different setting. Writing should be super careful, or readers get confused **)
 We also note that the usual semantics could indicate $u_2$ is the best if the scale of $\varphi_1$ were smaller enough.
 For example, in the case that the value are given by ''rotations per hour'', the robustness $\sem{\Model(u_2), \varphi_1} = -0.15$ is greater than $\sem{\Model(u_1), \varphi_3} = -2$ and consequently
 $\sem{\Model(u_2), \ATBRKIN} > \sem{\Model(u_1), \ATBRKIN}$.
 This illustrates a concrete example of the \emph{scale problem}.
\end{auxproof}

\end{myexample}

\subsection{Integrating MCR with Optimization-based Synthesis}
% \todo[inline]{improve}
% Integrating MCR with Algorithm~\ref{algo:SynthesisByCmaes} is realized by the following modification.
% The difference between conjunctive synthesis by MCR (Algorithm~\ref{algo:conjunctiveSynthesisByMCR}) and the original one is essentially the line~\ref{line:mcrSort}, where individuals are sorted by value of $F_X$ instead of their robustness.
% Incidentally robustness $\sem{\Model(u_i), \varphi_j}$ for each conjuncts $\varphi_1,\dots,\varphi_m$ is obtained explicitly at the line~\ref{line:conjunctRob}.
% This shows MCR works as a wrapper of the original fitness function, namely $u \mapsto \sem{\Model(u), \varphi_1 \land \dots \land \varphi_m}$, suggesting modularity of implementation.
% Note that this can be done only if all individuals in population are available, namely ask-and-tell interface is provided by the evolutionary algorithm.

% \todo[inline]{next version}
% \mw{The pseudocode should be summarized to clarify the difference between Algorithm~\ref{algo:SynthesisByCmaes} and Algorithm~\ref{algo:conjunctiveSynthesisByMCR}.}
Integrating MCR with Algorithm~\ref{algo:SynthesisByCmaes} is easy from the implementation point of view.
The resulting algorithm is showed as Algorithm~\ref{algo:conjunctiveSynthesisByMCR}.
We note that  $F_X$ at Line~\ref{line:evalMCR} implicitly depends on the values of robustness $\Rob_i^j\quad(i=1,\dots,\lambda, j=1,\dots,m)$ according to the definition of (\ref{eq:rankingFunction}).
Between the resulting algorithm and the original one, the only essential difference is in Line~\ref{line:mcrSort}---Algorithm~\ref{algo:conjunctiveSynthesisByMCR} uses the MCR rank $F_{\Pop}$. Therefore we expect that integrating MCR with other evolutionary computation-based falsification solvers can be done in a modular manner, too.

\begin{algorithm}[hb]
\caption{Conjunctive Synthesis by MCR}
\label{algo:conjunctiveSynthesisByMCR}
\begin{algorithmic}[1]
\Require Same as Algorithm~\ref{algo:SynthesisByCmaes}
\Function{Mcr-Synthesize}{$\Model,\varphi, \lambda, \theta^0$}
%\State $\vec \bu \gets \vec \bu_0$
\State $\Rob \gets -\infty$\;; \quad $u \gets \bot$
\State $g \gets 0$
\While{$\left(\parbox[]{4cm}{$\Rob \le 0$, within budget, and not stationary}\right)$}
  \State $X := \Pop \gets \textsc{Ask}(\theta^g)$
  \Statex  \Comment{Sample population}

  \For{$i \gets 1$ to $\lambda$}
    \State $u_i \gets \textsc{Gen-Signal}(\Indiv_i)$
    \State $\Rob_i^1 \gets \sem{\Model(u_i), \varphi_1}$
    \For{$j \gets 2$ to $m$}
      \State $\Rob_i^j \gets \sem{\Model(u_i), \varphi_j}$ \label{line:conjunctRob}
    \EndFor
    \State $\Rob_i \gets \min(\Rob_i^1, \Rob_i^2, \dots, \Rob_i^m)$
    \IIf{$\Rob_i > \Rob$}
      $\Rob \gets \Rob_{i}$\;; $u \gets u_i$ \EndIIf
    \Statex \Comment {Update robustness and signal}
  \EndFor
  \State $F_1, \dots, F_\lambda \gets 
    F_X(u_1), \dots F_X(u_\lambda)$ \label{line:evalMCR}
%    \langle\Rob_1^j\rangle_{j=1,\dots,m}, \dots,
%    \langle\Rob_\lambda^j\rangle_{j=1,\dots,m})$ 
  \State \parbox[t]{7cm}{$\Indiv_1, \dots \Indiv_\lambda \gets \Indiv_{s(1)}, \dots, \Indiv_{s(\lambda)}$ \newline
       with $s(i) \Defeq \text{argsort}(\langle F_1, \dots, F_\lambda\rangle, i)$} \label{line:mcrSort}
    \Statex \Comment {Sort individuals in ascending order by $F_X(x)$}
	\State $\theta^{g+1}\gets \textsc{Tell}(\theta^g, \Pop)$
      \Statex \Comment {Update parameter}
  \State $g\gets g+1$
\EndWhile%{$R\ge 0$ and within the budget}

\vspace{.3em}
\State$u\gets
\begin{cases}
 u & \text{if $\Rob>0$, that is, $\Rob=\sem{\Model(u), \varphi}>0$} \\
 \text{Failure}  & \text{no satisfying input found}
\end{cases}
$
\State \Return{$u$}
\EndFunction
\end{algorithmic}
\end{algorithm}

%####################
\section{Experiments}\label{sec:expr}
%####################

We implemented our conjunctive synthesis algorithm (denoted by ``MCR'') by combining Breach (\cite{Donze10}) with MCR.\@
In our implementation, we replaced the MATLAB implementation of CMA-ES with \emph{pycma} (a standard Python implementation of CMA-ES by~\cite{Hansen2005}) and combined with MCR (also implemented in Python).
%We combined Breach with the Python implemenation using the Python interface of MATLAB.\@

We conducted experiments to answer the following research questions.

\begin{description}
 \item[RQ1] Does MCR successfully address  the scale problem?
 \item[RQ2] How much does the choice of the objective conjunct ($\varphi_{1}$ in Problem~\ref{prob:ConjSynthByMCO}) influence the performance of MCR?
 \item[RQ3] Does MCR incur critical overheads?
\end{description}

\subsection{Benchmark Models and Specifications}%\mw{We should not use the same level as each benchmark (e.g. AT) because they are logically in a different level. Maybe we should use subsection instead.}
In our experiments, we used the following  benchmark models and specifications. The Simulink models are widely used in the benchmarks for falsification; see e.g.~\cite{ARCH20}. Each of the specifications consists of two to three conjuncts (cf.\ Problem~\ref{prob:ConjSynth}).
%Our experiments used three Simulink models that were selected as benchmarks for falsification in~\cite{ARCH20}.\mw{``We'' use the Simulink models ``in the experiments'' (subject is not the experiments)}

\paragraph*{Automatic Transmission (AT)}
The benchmark model AT was proposed in~\cite{ARCH15:Benchmarks_for_Temporal_Logic}.
This model has two inputs \textsf{throttle} and \textsf{break} each with 5 control points; and has three outputs $\mathsf{gear} \in \{0,\dots,3\}$, \textsf{rpm} and \textsf{speed}.
% The timed output signals only depend on inputs.
Our goal is to find a satisfying signal for each of the following specifications.
%The specifications we try to satisfy are following:\mw{We should complete the sentence e.g. ``Our goal is to find a satisfying signal for each of the following specifications.''}
\begin{align*}
  \ATBRKIN_p \DefSpec\; & \ATBRKINca \land \ATBRKINcb \\ &\land \ATBRKINcc \\
  \ATSpeed \DefSpec\; & \ATSpeedFormula \\
  \ATSpeedRpm_{p_1,p_2}  \DefSpec\; & \ATSpeedRpmFormula
\end{align*}
While $\mathsf{gear} \in \{1,2,3\}$, we  have $\mathsf{RPM} \in [0, 5000]$ and $\mathsf{speed} \in [0,100]$, and thus the scale problem is expected  in $\ATBRKIN_p$ and $\ATSpeedRpm_{p_1,p_2}$.
%At the time gear takes value $\mathsf{gear} \in \{1,2,3\}$, \textsf{RPM} can reach 5000 and \textsf{speed} usually stays between 0 and 100.\mw{``can reach'': do you mean at most 5000? Do you mean the following? ``When $\mathsf{gear} \in \{1,2,3\}$, we typically have $\mathsf{RPM} \in [0, 5000]$ and $\mathsf{speed} \in [0,100]$.''}

\paragraph*{Fuel Control (AFC)}
%The model is described in [19] and has been used in two previous instalments of this competition [7, 8]. The specific limits used in the requirements are chosen such that falsification is possible but reasonably hard.
%The constrained input signal (instance 2) fixes the throttle ホク to be piecewise constant with 10 uniform segments over a time horizon of 0 with two modes (normal and power corresponding to feedback and feedforward control), and the engine speed マ‰ to be constant with 900 竕、 マ‰ < 1100 to capture the input profile outlined in [19] and to match the previous competitions. For this reason, we do not consider the unconstrained (instance 1) input specification. Faults are disabled (e.g. by setting fault_time > 50).
The benchmark model AFC was proposed in~\cite{JinDKUB14}.
This model has two inputs \textsf{throttle} and \textsf{engine}, each with 10 control points.
It has two outputs $\mathsf{mode}$ and $\mu$: $\mathsf{mode} \in \{0,1\}$ shows if the system is controlled by feedback control or feedforward control; and $\mu$ is the deviation of the air-fuel ratio from its reference value.
We aim to find a signal that satisfies the following specification.
\begin{align*}
  \AFCSpec \DefSpec\; & \AFCca \land \AFCcb \\
    & \land \AFCcc \\
    & \land \AFCcd
\end{align*}

\paragraph*{Wind Turbine (WT)}
%Wind Turbine (WT). The model is a simplified wind turbine model proposed in [25]. The input of the system is wind speed v and the outputs are blade pitch angle ホク, generator torque Mg,d, rotor speed ホゥ and demanded blade pitch angle ホクd. The wind speed is constrained by 8.0 竕、 v 竕、 16.0. Instance 1 allows any piece-wise continuous inputs, while instance 2 constrains inputs to piece-wise constant signals whose control points which are evenly spaced each 5 seconds. The model is relatively large. Further, the time horizon is long (630) compared to other benchmarks.
The benchmark model WT was proposed in~\cite{ARCH16:Hybrid_Modelling_of_Wind}.
This model has one input $v$ with 18 control points, and has three outputs that are the blade pitch angle $\theta$, the generator torque $M_{g,d}$, and the rotor speed $\Omega$.
% The timed output signals only depend on inputs.
Our goal is to find a signal that satisfies the following specification.
\begin{align*}
  \WTSpec \DefSpec\; &  \WTca \\
    & \land \WTcb \land \WTcc
\end{align*}
We used the time horizon of  90 seconds while it is 630 seconds in~\cite{ARCH20}.
We observed that 90 seconds is long enough to satisfy the specification \WTSpec.
% Though WT benchmark in~\cite{ARCH20} is conducted under 630 seconds of simulation time,
% we cut off the simulation time to 90, which is enough to observe the falsification of the specification \WTSpec.
% \mw{Do you mean, we modified the time horizon of the input/output signals?}

\subsection{Experiment}

We compared the performance of our algorithm (henceforth denoted by ``MCR'') with the two baseline implementations ``Breach'' and ``Breach\_pycma.'' 
Although they are algorithmically the same, we used both of them 
%as the baselines because their solvers 
since they have some minor differences, e.g., in the handling of input ranges. 
% As there are differences in the details of the implementation of CMA-ES solver like 
% we regard both Breach and Breach\_pycma as our baselines.
\begin{itemize}
 \item (``Breach'') Breach 1.7.0 with its original MATLAB implementation of CMA-ES % as the optimization solver; and
 \item (``Breach\_pycma'') Breach 1.7.0 with its CMA-ES implementation replaced with the one offered by the  pycma library in Python~\cite{Hansen2005}
       % implementation of CMA-ES in Python  %$$as the optimization solver 
\end{itemize}
Note that these baselines do not include MCR as a constraint-handling technique.

We conducted the experiments on an Amazon EC2 c4.xlarge instance (2.9 GHz Intel Xeon E5-2666 v3, 7.5 GB RAM).\@

For each problem instance, we executed each of the implementations for 20 times with different initial seeds.
%Our experiment consists in the execution of Breach, Breach\_pycma and MCR over our benchmarks for 20 trials, using different initial seeds.
We set a timeout in 600 seconds. %, which is long enough to compare success rates.
We measured the \emph{success rate} (SR) that is the number of the successful trials, i.e., the trials where a signal satisfying the given specification was found. We also measured the \emph{average elapsed time} of the successful trials.

As we mentioned in Remark~\ref{rem:choiceOfObjective}, when one translates optimization-based falsification into constrained optimization, there is freedom in the choice of the objective conjunct ($\varphi_{1}$ in Problem~\ref{prob:ConjSynthByMCO}).
 % conjunct as the optimization target ($\varphi_{1}$ in Problem~\ref{prob:ConjSynthByMCO}).
% When we try to solve Problem~\ref{prob:ConjSynthByMCO} by MCR, the choice of objective function indeed affects the behavior of the optimization.
This choice affects the performance of MCR.
In our experiments, we tried each conjunct in a specification as the optimization target, and we report the performance of the best and the worst choices.

Table~\ref{tab:expResults} summarizes the experiments results. We mark time as ``---'' if SR is 0, i.e., if all trials failed.

\newcommand{\best}{\cellcolor{green!25}}
\newcommand{\mf}[1]{\textbf{\color{blue} #1}}

\begin{table*}[tbp]
\caption{Experimental results. SR shows success rates (out of 20 trials); time is the average execution time for successful trials in seconds. MCR (best) (resp.\ MCR (worst)) represents the best (resp.\ worst) instance of MCR, in terms of which conjunct we chose as the objective $\varphi_{1}$ (cf.\ Problem~\ref{prob:ConjSynthByMCO}). 
For each problem instance, the best combination in terms of the following order is highlighted: $(\mathrm{SR}, \mathrm{time})$ is better than $(\mathrm{SR}', \mathrm{time}')$ if and only if we have $\mathrm{SR}>\mathrm{SR'}$ or we have both $\mathrm{SR}=\mathrm{SR}'$ and $\mathrm{time} < \mathrm{time}'$.
For each instance, the largest SR is shown in blue.
% the column ``objective'' is the formula of conjunct which the instance chosen as the objective.
}\label{tab:expResults}
\centering\scriptsize
\begin{tabular}{lrrcrrcrrcrrrc}
\toprule
\multirow{2}{*}{Spec.\ $\varphi$} & \multicolumn{2}{c}{Breach} && \multicolumn{2}{c}{Breach\_pycma} && \multicolumn{3}{c}{MCR (best)} && \multicolumn{3}{c}{MCR (worst)} \\
\cmidrule{2-3} \cmidrule{5-6} \cmidrule{8-10} \cmidrule{12-14}
& SR (/20) & time [s] && SR & time && SR & time & objective conjunct $\varphi_{1}$ && SR & time & objective conjunct  $\varphi_{1}$ \\
\midrule
$\ATBRKIN_{2500}$       &14 & 27.3 && \mf{20} & 47.9 &&  \best \mf{20} & \best35.6 &  $\ATBRKINcc$ && \mf{20} & 101.7 & $\ATBRKINca$ \\
$\ATBRKIN_{2400}$       & 4 & 36.4 &&  7 & 69.3 &&  \best \mf{19} & \best 188.2 &  $\ATBRKINca$ &&  6 & 96.4 &  $\ATBRKINcc$\\
$\ATBRKIN_{2300}$       & 0 & ---  &&  0 &  --- &&  \best \mf{9} & \best 349.4 & $\ATBRKINca$&&   0 & --- & $\ATBRKINcc$ \\
$\ATSpeed$              & 12& 190.8&& \best \mf{16} & \best 298.0&&  15 & 340.1 & $\ATSpeedca$ && 13 & 194.1 &  $\ATSpeedcb$ \\
$\ATSpeedRpm_{80,4500}$ & \mf{20} & 38.6&& \mf{20} &  59.3 && \best \mf{20} & \best 30.9  & $\ATSpeedRpmcb$ &&  \mf{20} & 43.2 & $\ATSpeedRpmca$ \\
$\ATSpeedRpm_{50,2700}$ & 19 & 95.2&& \mf{20} & 241.4 && \best \mf{20} & \best 237.5 & $\ATSpeedRpmca$ && 16 & 254.5 & $\ATSpeedRpmcb$ \\
\midrule
$\AFCSpec$                    & 4 & 325.3 && 7 & 194.0 && \best \mf{14} & \best 229.8 & $\AFCcb$ && 10 & 230.6 &
% $\AFCcc$ 
\parbox[t]{3cm}{$\BoxOp{[0,30]}(\mathsf{throttle} > 40 \Rightarrow \mathsf{engine} < 1000)$}
\\
\midrule
$\WTSpec$                     & 10 & 327.9 && 16 & 240.4 && \best \mf{20} & \best 140.7 & $\WTcc$ && 19 & 162.4  & $\WTcb$ \\
\bottomrule
\end{tabular}
\end{table*}

% \paragraph{Comparison}
% \todo[inline]{Comparison needed? possibly time-SR step graph}
% \mw{We can omit this subsection and directly have a subsection for each RQ}
\subsection{Discussion}
% \todo[inline]{extend}
% \todo[inline]{Should we introduce research questions in an early part of this section?}
% We see the results with some research questions.
We discuss the three research questions in view of Table~\ref{tab:expResults}.

\researchquestion{Does MCR successfully address  the scale problem?}\label{rq:scaleProblem}

The experiment results 
%in Table~\ref{tab:expResults} 
give an affirmative answer. The advantage of MCR is more obvious in challenging problem instances such as  $\ATBRKIN_{p}$, AFC, and WT. AT2 and AT3 are less challenging ones where the scale problem is less eminent; for these problem instances, too, MCR's performance is comparable or better compared to plain Breach.

Note that $\ATBRKIN_{p}$ with $p=2400$ is the same specification that we discussed in Example~\ref{ex:AT1withMCR}; we discussed there that the specification is subject to the scale problem. Note also that the problem becomes harder as the parameter $p$ becomes smaller. 

In Table~\ref{tab:expResults}, we see that Breach (both with the original and pycma CMA-ES) suffer from the scale problem; notably it succeeds in zero out of 20 trials for the hardest instance ($\ATBRKIN_{2300}$). In those unsuccessful runs of Breach, the evolutionary computation tends to  terminate because of stationarity (Line~\ref{line:whileGuard} of Algorithm~\ref{algo:SynthesisByCmaes}), not because of timeout. This is because the robustness of $\varphi_{3}$ (regarding $\textsf{gear}$) becomes dominant at some stage but $\varphi_{3}$'s robustness is hard to improve (cf.\ the masking effect in the scale problem, Example~\ref{ex:AT1withMCR}). 

In contrast, MCR often succeeds to synthesize a satisfying input---almost half the time even for  the hardest instance $\ATBRKIN_{2300}$.

% Yes. For the benchmarks \ATBRKIN, \AFCSpec and \WTSpec, which exhibit the scale problems and consequently are considered as a rather difficult problem to satisfy, the best instance of MCR apparently outperforms the existing ones in SR.\mw{We should refer each specification so that the readers notice that the existence of the scale problem is clear purely from the specification (without any experiments).}

% Especially for AT1, it is clear that the combination with MCR is advantageous in solving our target problem (the conjunctive synthesis problem, Problem~\ref{prob:ConjSynth}), especially for its harder instances $\ATBRKIN_{2300}$.
% This is because, in trials of AT1 with Breach or Breach\_pycma, the best robustness in the process of evolution tends to stay in constant $-1$ or $-2$, which exhibits the scale problem we illustrated in Example~\ref{ex:AT1withMCR}; the robustness values other than one concerning \textsf{gear} is masked. Consequently, quite a few trials terminated unsuccessfully before the timeout, indicating a terminal criterion like \texttt{EqualFunValues} that we mentioned in \S{}\ref{sec:synthesisByCmaes} is fulfilled.

% On the other hand, for easy benchmarks like AT2 and AT3, it is unclear whether the scale problem is avoided. However, for $\ATSpeedRpm_{80,4500}$, the elapsed time was improved for good instances of MCR. 

\researchquestion{
How important is the choice of the objective conjunct in MCR?
%Does the choice of objective have a critical impact on performance?
}\label{rq:choiceOfObjective}

The problem instance in which the performance gap between MCR (best) and MCR (worst) is the largest is $\ATBRKIN_{2300}$, the apparently hardest instance in Table~\ref{tab:expResults}.
Apart from that, the impact of the choice of the objective is seen to be moderate. We also observe that the performance of MCR (worst) is comparable or better compared to Breach without MCR.

\researchquestion{Does MCR incur critical overheads?}\label{rq:overheads}

The comparison between Breach and MCR in the execution time is mixed in Table~\ref{tab:expResults}. MCR takes longer for $\ATBRKIN_{2400}$, for example, but at the same time it achieves much higher success rate. For problem instances such as $\ATSpeedRpm_{80,4500}$ and $\WTSpec$, MCR finishes earlier while its success rate is the same as or higher than Breach. 

In Table~\ref{tab:expResults},  the average execution time of MCR is only several times longer than than that of Breach, even in the worst instance. Therefore we conclude that the overhead of MCR is admissible, in view of its higher success rate.

% For AT1, the elapsed time of MCR is more costly than the baseline approaches.
% As the baseline approaches tend to fail in trials, it is not fair to conclude that MCR has low effectiveness from these results.

% For AT2, in which both the baselines and our approach are almost equivalently successive, the elapsed time improved (for MCR (best)) or not much different (for MCR (worst)). For $\ATSpeedRpm_{50,2700}$, contrariwise, MCR took much low more time to find satisfying input than the baseline approaches. Therefore overheads highly depend on the nature of models and specifications.

\section{Conclusions and Future Work}\label{sec:conclusion}

In this paper, we propose a view of the conjunctive synthesis problem as a constrained optimization problem.
This approach enables solvers to use constraint handling techniques i.e., MCR.\@
Our experiments show that the approach remarkably improves the success rate of synthesis when the scale problem involves.

One future work is to extend our idea of using the constrained optimization problem to a more general form of specifications than the disjunctive specifications in the synthesis problem. Investigating a  method to choose a good objective conjunct is another future work.

\begin{ack}
The authors are supported by ERATO HASUO Metamathematics for Systems Design Project (No. JPMJER1603), JST.\@
\end{ack}

\bibliography{export}             % bib file to produce the bibliography
                                                     % with bibtex (preferred)

\end{document}